# Reconstruction of unstable heavy particles using deep symmetry-preserving attention networks

Check for updates

Michael James Fenton ®[1,7] ✉, Alexander Shmakov[2,7] ✉, Hideki Okawa ®[3], Yuji Li[4], Ko-Yang Hsiao[5], Shih-Chieh Hsu ®[6], Daniel Whiteson ®[1] & Pierre Baldi ®[2]

Reconstructing unstable heavy particles requires sophisticated techniques to sift through the large number of possible permutations for assignment of detector objects to the underlying partons. An approach based on a generalized attention mechanism, symmetry preserving attention networks (SPA-NET), has been previously applied to top quark pair decays at the Large Hadron Collider which produce only hadronic jets. Here we extend the SPA-NET architecture to consider multiple input object types, such as leptons, as well as global event features, such as the missing transverse momentum. In addition, we provide regression and classification outputs to supplement the parton assignment. We explore the performance of the extended capability of SPA-NET in the context of semi-leptonic decays of top quark pairs as well as top quark pairs produced in association with a Higgs boson. We find significant improvements in the power of three representative studies: a search for $t\bar{t}H$, a measurement of the top quark mass, and a search for a heavy $Z'$ decaying to top quark pairs. We present ablation studies to provide insight on what the network has learned in each case.

Event reconstruction is a crucial problem at the Large Hadron Collider (LHC), where heavy, unstable particles such as top quarks, Higgs bosons, and electroweak $W$ and $Z$ bosons decay before being directly measured by the detectors. Measuring the properties of these particles requires reconstructing their four-momenta from their immediate decay products, which we refer to as *partons*. Since many partons leave indistinguishable signatures in detectors, a central difficulty is assigning the observed detector objects to each parton. As the number of partons grows, the combinatorics of the problem becomes overwhelming, and the inability to efficiently select the correct assignment dilutes valuable information.

Previously, methods such as $\chi^2$ fits[1] or kinematic likelihoods[2] have provided analytic approaches for performing this task. These approaches are limited, however, by the requirement of exhaustively building each possible permutation of the event and by the limited amount of kinematic information that can be incorporated. Particularly at high-energy hadron colliders such as the LHC, events often contain many extra objects from additional activity as well as the particles originating from the hard scattering event, which can cause the performance of permutation-based methods to degrade substantially.

In recent years, modern machine learning tools such as graph neural networks and transformers[3] have been broadly applied to many problems in high-energy physics. For example, the problem of identifying the origin of single, large-radius jets has been closely studied[4–14] using such techniques. Some of these have incorporated symmetry considerations[11,12,14] to aid performance. Implementations of such strategies to event-level reconstruction have been limited so far to single object permutation assignment[15–17] or direct regression[18].

This work presents a complete machine learning approach to multi-object event reconstruction and kinematic regression at the LHC, named SPA-NET owing to its use of a symmetry-preserving attention mechanism, designed to incorporate all of the symmetries present in the problem. It was first introduced[15,16] in the context of reconstruction of the all-hadronic final state in which only one type of object is present. In this work, we extend and complete the method by generalizing to arbitrary numbers of object types, as well as adding multiple capabilities that can aid the application of SPA-NET in LHC data analysis, including signal and background discrimination, kinematic regression, and auxiliary outputs to separate different kinds of events.

[1]Department of Physics and Astronomy, University of California, Irvine, Irvine 92607 CA, USA. [2]Department of Computer Science, University of California, Irvine, Irvine 92607 CA, USA. [3]Institute of High Energy Physics, Chinese Academy of Sciences, Shijingshan 100049 Beijing, China. [4]Institute of Modern Physics, Fudan University, Yangpu 200433 Shanghai, China. [5]Department of Physics, National Tsing Hua University, Hsinchu City 30013, Taiwan. [6]Department of Physics and Astronomy, University of Washington, Seattle 98195-4550 WA, USA. [7]These authors contributed equally: Michael James Fenton, Alexander Shmakov.
✉e-mail: mjfenton@uci.edu; ashmakov@uci.edu

To demonstrate the new capacity of the technique, we study its performance in final states containing a lepton and a neutrino. The method is compared to existing baseline approaches and demonstrated to provide significant improvements in three flagship LHC physics measurements: $t\bar{t}H$ cross-section, top quark mass, and a search for a hypothetical $Z'$ boson decaying to top quark pairs. These examples demonstrate various additional features, such as kinematic regression and signal versus background discrimination. The method can be applied to any final state at the LHC or other particle collider experiments, and may be applicable to other set assignment tasks in other scientific fields.

## Methods
### SPA-NET extensions
We present several improvements to the base SPA-NET architecture[15,16] to tackle the additional challenges inherent to events containing multiple reconstructed object classes and to allow for a greater variety of outputs for an array of potential auxiliary tasks. These modifications allow SPA-NET to be applied to essentially any topology and allow for the analysis of many additional aspects of events beyond the original jet-parton assignment task.

**Base SPA-NET overview.** For context, we first provide a brief overview of the original SPA-NET architecture[15,16]. These components are those which are presented with black boxes and lines in Fig. 1. The jets, represented by their kinematics, are first embedded into a high dimensional latent space and subsequently processed by a central transformer encoder[3] with the goal of providing contextual information to the jets. We note that the architecture of this transformer encoder follows the original definition[3], with one major exception: we omit the positional encoding to prevent introducing ordering over our input. As the jets are presented as a set of momentum vectors, with no obvious order, we want the network to remain permutation equivariant with respect to the input order. We replicate the architecture for the particle transformers, now applying individually trained transformers for every resonance particle in our event.

Finally, to extract the joint distribution over jets for each resonance particle, we apply a symmetric tensor attention layer defined in Section 3 of our previous work[16]. This layer applies a generalized form of attention, modified by a symmetry group over assignments, to produce a symmetric joint distribution over jets describing the likelihood of assigning said jets to the resonance particle. This split architecture, with individual branches for every resonance particle, allows us to avoid computing a full permutation over all possible assignments and reduced the runtime from combinatorial w.r.t the number of jets, $\mathcal{O}(N!)$, to $\mathcal{O}(N^{k_p})$ where $k_p$ is the number of daughter particles produced by a resonance particle.

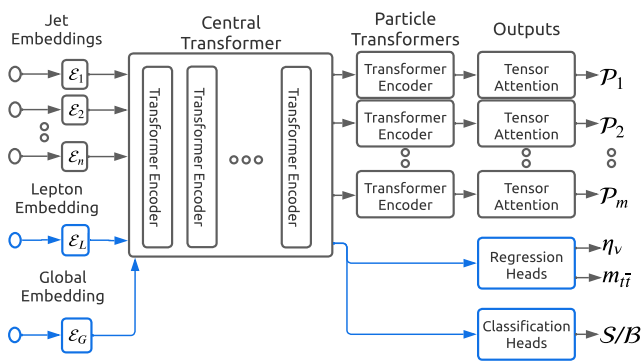

**Fig. 1 | Extended diagram of the new SPA-NET architecture.** The diagram flows left to right, with inputs denoted by $\mathcal{E}_i$, assignment outputs denoted by $P_j$, regression outputs $\eta_\nu$ and $m_{t\bar{t}}$, and classification output $S/B$. Black blocks show components common to our previous works[15,16], with new components shown in blue.

**Input observables.** While the original SPA-NET[15,16] studies concentrated on examples where all objects have hadronic origins, we focus here on the challenges of semi-leptonic topologies. These events contain several different reconstructed objects, including the typical hadronic jets as well as leptons and missing transverse momentum ($E_T^{miss}$) typically associated with neutrinos. Unlike jets or leptons, this $E_T^{miss}$ is a global observable, and its multiplicity does not vary event by event.

We accommodate these additional inputs by training individual position-independent embeddings for each class of input. This allows the network to adjust to the various distributions for each input type, and allows us to define sets of features specific to each type of object. We parameterize jets using the $\{M, p_T, \eta, \sin\phi, \cos\phi, b-\text{tag}\}$ representation, where $M$ is the jet mass, $p_T$ is the jet momentum transverse to the incoming proton beams, and $\phi$ is the azimuthal angle around the detector, represented by its trigonometric components to avoid the boundary condition at $\phi = \pm \pi$. $\eta$ is the pseudo-rapidity[19] of the jet, the standard measure of the polar angle between the incoming proton beam and the jet commonly used in particle physics due to its Lorentz-invariant quantities. Leptons are similarly represented using $\{M, p_T, \eta, \sin\phi, \cos\phi, \text{flavor}\}$ where flavor is 0 for electrons and 1 for muons. Finally, $E_T^{miss}$ is represented using two scalar values, the magnitude and azimuthal angle, and is treated as an always-present jet or lepton. The individual embedding layers map these disparate objects with different features into a unified latent space which may be processed by the central transformer.

The global inputs, such as $E_T^{miss}$, need to be treated differently than the jets and leptons, as they do not have associated parton assignments. Therefore, after computing the central transformer, we do not include the extra global $E_T^{miss}$ vector in the particle transformers. This allows the transformer to freely share the $E_T^{miss}$ information with the other objects during the central transformer step while preventing it from being chosen as a reconstruction object for jet-parton assignment.

**Secondary outputs.** Beyond jet-parton assignment, we are interested in reconstruction of further observables, such as the unknown neutrino $\eta$, or differentiation of signal events from background. These observables are defined at event level, and are independent of the jet multiplicity, so we must construct a way of summarizing the entire event in a single vector to predict these values.

To accomplish this, we add additional output heads to the central transformer, presented with blue boxes and lines on the right in Fig. 1, which are trained end-to-end simultaneously with the base reconstruction task. We extract an event embedding from the central transformer by including a learnable event vector in the inputs to the transformer. We append this learned event vector $\mathcal{E}_E \in \mathbb{R}^D$ to the list of embedded input vectors: $\mathcal{E} = \{\mathcal{E}_1, \mathcal{E}_2, \ldots, \mathcal{E}_n, \mathcal{E}_L, \mathcal{E}_G, \mathcal{E}_E\}$ prior to the central transformer (Fig. 1). This allows the central transformer to process this event vector using all of the information available in the observables.

We extract the encoded event vector after the central transformer and treat it as a latent summary representation of the entire event $z_E$. We can then feed these latent features into simple feed-forward neural networks to perform signal vs background classification, $\mathcal{S}/\mathcal{B}(z_E)$, neutrino kinematics regression, $\eta_\nu(z_E)$, or any other downstream tasks. These tasks may additionally be learned after the main SPA-NET training as $z_E$ may be computed used a fixed set of SPA-NET weights and then used for other downstream tasks without altering the original SPA-NET.

These additional feed-forward networks are trained using their respective loss, either categorical log-likelihood or mean squared error (MSE). These auxiliary losses are simply added to the total SPA-NET loss, weighted by their respective hyperparameter $\alpha_i$. With the parton reconstruction loss, $\mathcal{L}_{\text{reconstruction}}$ defined as the masked minimum permutation loss from Equation 6 of our previous work[16], the SPA-NET loss becomes:

$$\mathcal{L} = \alpha_{\text{reco}} \mathcal{L}_{\text{reconstruction}} + \alpha_{\text{clas}} \mathcal{L}_{\text{classification}} + \alpha_{\text{regr}} \mathcal{L}_{\text{regression}}. \quad (1)$$

**Particle detector.** In our previous work[16], we introduced the ability to reconstruct partial events by splitting the reconstruction task based on the event topology. This is a powerful technique that is particularly useful in complex events, where it is very likely that at least one of the partons will not have a corresponding detector object.

However, the assignment outputs are trained only on examples in which the event contains all detector objects necessary for a correct parton assignment. We refer to the reconstruction target particles in these examples as reconstructable. We must train this way because only reconstructable particles have truth-labeled detector objects, which are required for training, and we ignore non-reconstructable particles via the masked loss defined in Equation 6 of our previous work[16]. As a result of this training procedure, the SPA-NET assignment probability $P_a$ only represents a conditional assignment distribution over jet indices $j_i$ for each particle $p$ given that the particle is reconstructable:

$$P_a(j_1, j_2, \ldots, j_{k_p} | p \text{ reconstructable}). \quad (2)$$

We use $P(p \text{ reconstructable}) = P(p)$ and $P(p \text{ not reconstructable}) = P(\neg p)$ for conciseness. To construct an unconditional assignment distribution, we need to additionally estimate the probability that a given particle is reconstructable in the event, $P_d$. This additional distribution may be used to produce a pseudo-marginal probability for the assignment. While $P_a(j_1, j_2, \ldots, j_{k_p} | \neg p) = 0$ is not a valid distribution, and therefore this marginal probability is ill-defined, we may still use this pseudo-marginal probability

$$\mathcal{P}(j_1, j_2, \ldots, j_{k_p}) = P_a(j_1, j_2, \ldots, j_{k_p} | p) P_d(p) \quad (3)$$

as an overall measurement of the assignment confidence of the network.

We aim to estimate this reconstruction probability, $P_d(p)$, with an additional output head of SPA-NET. We will refer to this output as the detection output, because it is trained to detect whether or not a particle is reconstructable in the event. We train this detection output in a similar manner as the classification outputs but at the particle level instead of the event level. That is, we extract a summary particle vector from each of the particle transformer encoders using the same method as the event summary vector from the central transformer. We then feed these particle vectors into a feed-forward binary classification network to produce a Bernoulli probability for each particle. We have to also take into account the potential event-level symmetries in a similar manner to the assignment reconstruction loss from Equation 6 of our previous work[16]. We train this detection output with a cross-entropy loss over the symmetric particle masks:

$$\mathcal{L}_{\text{detection}} = \min_{\sigma \in G_E} \left[ \mathcal{M}_{\sigma(p)} \log P_d(p) + (1 - \mathcal{M}_{\sigma(p)}) \log(1 - P_d(p)) \right]. \quad (4)$$

The complete loss equation for the entire network can now be defined:

$$\mathcal{L} = \alpha_{\text{reco}} \mathcal{L}_{\text{reconstruction}} + \alpha_{\text{det}} \mathcal{L}_{\text{detection}} + \alpha_{\text{clas}} \mathcal{L}_{\text{classification}} + \alpha_{\text{regr}} \mathcal{L}_{\text{regression}}. \quad (5)$$

**Baseline methods**

We compare SPA-NET to two commonly used methods, the Kinematic Likelihood Fitter (KLFitter)[2], and a Permutation Deep Neural Network (PDNN), which uses a fully connected deep neural network similar to existing literature[20]. Both methods are permutation-based, meaning they sequentially evaluate every possible permutation of particle assignments. This results in a combinatorial explosion, with for example $5!/2 = 60$ possible assignments of the jets in a semi-leptonically decaying $t\bar{t}$ + jet event (the reduction by a factor of two comes from the assignment symmetry between the hadronically decaying $W$ boson decay products). That is, there are 60 different possible permutations that must be evaluated per event, even before considering systematic uncertainty evaluation or further additional jets. With typical analyses utilizing MC samples containing $\mathcal{O}(10^6 - 10^8)$ events, which must be evaluated for $\mathcal{O}(10^2)$ systematic variations, complex events quickly become intractable or at least extremely computationally expensive, even before considering the decreasing performance of such methods as a function of object multiplicity. The performance of these algorithms is compared to SPA-NET in all presented results.

**KLFitter.** KLFitter has been extensively used in top quark analyses[21-29], especially for semi-leptonic $t\bar{t}$ events. The method involves building every possible permutation of the event and constructing a likelihood score for each. The permutation with the maximum likelihood is thus taken as the best reconstruction for that event. The likelihood score, which has been updated (https://github.com/KLFitter/KLFitter) since the original publication[2], is defined as

$$\begin{aligned}
\mathcal{L} &= B\left(m_{q_1 q_2 q_3} | m_t, \Gamma_t\right) \cdot B\left(m_{q_1 q_2} | m_W, \Gamma_W\right) \\
&\times B\left(m_{q_4 \ell \nu} | m_t, \Gamma_t\right) \cdot B\left(m_{\ell \nu} | m_W, \Gamma_W\right) \\
&\times \prod_{i=1}^{4} W_{jet}\left(E_{jet,i}^{meas} | E_{jet,i}\right) \cdot W_\ell\left(E_\ell^{meas} | E_\ell\right) \\
&\times W_{miss}\left(E_x^{miss} | p_x^\nu\right) \cdot W_{miss}\left(E_y^{miss} | p_y^\nu\right),
\end{aligned} \quad (6)$$

where $B$ represents Breit-Wigner functions, $m_{q_1 q_2 q_3}$, $m_{q_1, q_2}$, $m_{q_4 \ell \nu}$, $m_{\ell \nu}$ are invariant masses computed from the final state particle momenta. The variables $m_{t(W)}$ and $\Gamma_{t(W)}$ are the masses and decay widths of the top quark ($W$ boson), respectively. The expressions $E_{\ell,jet}^{(meas)}$ represents the (measured) energy of the leptons or jets, respectively, and the functions $W_{var}(var_A | var_B)$ are the transfer function for the variable $var_A$ from $var_B$.

This method suffers from several limitations. Firstly, the requirement to construct and test every possible permutation leads to a run-time that grows exponentially with the number of jets or other objects in the event. This quickly becomes a limiting factor in large datasets, which at the LHC often contain millions of events that must be evaluated hundreds of times each (once per systematic uncertainty shift). While semi-leptonic $t\bar{t}$ can largely remain tractable, it can significantly slow down analyses due to the heavy computing cost, and it is typical to limit the evaluation to only a subset of the reconstructed objects in order to reduce this burden, which restricts the number of events that can be correctly reconstructed. More complex final states, for example $t\bar{t}H$ production, require even more objects to be reconstructed and thus take even longer to compute, severely limiting the usability of the method in such channels.

A second limitation of the method is its treatment of partial events, which the likelihood is not designed to handle, and thus performance in these events is significantly degraded. Finally, the method does not take into account any correlations between the decay products of the target particles and the rest of the event, since only the particles hypothesized as originating from the targets are included in the likelihood evaluation. An advantage of the method is the use of transfer functions to represent detector effects, but these must be carefully derived for each detector to achieve maximum performance, which can be a difficult and time-consuming endeavor.

There are two variations of the KLFitter likelihood of interest in our studies: one in which the top quark mass is given an assumed value, and one in which it is not. Specifying the assumed mass leads to improved reconstruction efficiency at the expense of biasing towards permutations at this mass, causing sculpting of backgrounds and other undesirable effects. In the analyses presented in $t\bar{t}H$ and $Z'$ analyses, the top quark mass is fixed to a value of 173 GeV, since this biasing is less important than overall reconstruction efficiency. In contrast, the top quark mass measurement must avoid biasing towards a specific mass value, and thus the mass is not fixed in the likelihood for this analysis.

**PDNN.** The PDNN uses a fully connected deep neural network that takes the kinematic and tagging information of the reconstructed objects as inputs, similar to the method described in existing literature[20]. Again, each possible permutation of the event is evaluated, and the assignment

Table 1 | Reconstruction efficiencies for hadronically decaying ($t_H$) and leptonically decaying ($t_L$) top quarks, Higgs bosons ($H$), and complete events (Ev.) for semi-leptonic $t\bar{t}$ and $t\bar{t}H(H \to b\bar{b})$ processes

| | $N_{jets}$ | SPANet Eff. (%) | | | | PDNN Eff. (%) | | | | KLFitter Eff. (%) | | | |
|---|---|---|---|---|---|---|---|---|---|---|---|---|---|
| | | Ev. | $t_H$ | $t_L$ | $H$ | Ev. | $t_H$ | $t_L$ | $H$ | Ev. | $t_H$ | $t_L$ | $H$ |
| All $t\bar{t}$ Events | =4 | 81 | 80 | 86 | – | 74 | 80 | 78 | – | 60 | 66 | 65 | – |
| | =5 | 74 | 72 | 84 | – | 68 | 69 | 79 | – | 32 | 37 | 47 | – |
| | ≥6 | 66 | 61 | 82 | – | 57 | 53 | 75 | – | 18 | 20 | 35 | – |
| | **All** | **76** | **73** | **85** | – | **68** | **69** | **78** | – | **42** | **44** | **53** | – |
| Full $t\bar{t}$ Events | =4 | 84 | 84 | 90 | – | 83 | 83 | 89 | – | 71 | 71 | 77 | – |
| | =5 | 73 | 74 | 87 | – | 69 | 71 | 84 | – | 28 | 39 | 52 | – |
| | ≥6 | 60 | 63 | 84 | – | 51 | 55 | 79 | – | 12 | 21 | 37 | – |
| | **All** | **75** | **76** | **87** | – | **70** | **72** | **85** | – | **41** | **47** | **58** | – |
| All $t\bar{t}H$ Events | =6 | 43 | 54 | 69 | 50 | 33 | 48 | 57 | 43 | 31 | 36 | 55 | 43 |
| | =7 | 39 | 48 | 68 | 49 | 31 | 43 | 58 | 43 | 16 | 25 | 47 | 29 |
| | ≥8 | 34 | 42 | 68 | 47 | 27 | 36 | 56 | 40 | 11 | 15 | 46 | 24 |
| | **All** | **40** | **49** | **69** | **49** | **31** | **43** | **57** | **43** | **21** | **26** | **50** | **34** |
| Full $t\bar{t}H$ Events | =6 | 54 | 65 | 73 | 62 | 49 | 60 | 68 | 58 | 38 | 49 | 60 | 48 |
| | =7 | 42 | 55 | 70 | 56 | 36 | 50 | 64 | 51 | 13 | 31 | 48 | 31 |
| | ≥8 | 33 | 47 | 69 | 52 | 28 | 42 | 60 | 46 | 05 | 17 | 46 | 24 |
| | **All** | **45** | **57** | **71** | **57** | **39** | **52** | **64** | **52** | **19** | **33** | **52** | **35** |

The efficiencies highlighted in bold are inclusive of jet multiplicity.

with the highest network output score is taken as the best reconstruction. Training is performed as a discrimination task, in which the correct permutations are marked as signal, and all of the other permutations are marked as background.

This method also suffers from several limitations, including the same exponentially growing run-time due to the permutation-based approach, the inability to adequately handle partial events, and the lack of inputs related to additional event activity. Further, the method does not incorporate the symmetries of the reconstruction problem due to the way in which input variables must be associated with the hypothesized targets. Recently, message-passing graph neural networks were applied to the all-hadronic $t\bar{t}$ final state[17], but as all studies presented here are performed in the lepton +jets channel, no comparison is made to such methods.

### Datasets and training
Several datasets of simulated collisions are generated to test a variety of experimental analyses and effects. All datasets are generated at a center-of-mass energy of $\sqrt{s} = 13$ TeV using MADGRAPH_AMC@NLO[30] (v3.2.0, NCSA license) for the matrix element calculation, PYTHIA8[31] (v8.2, GPL-2) for the parton showering and hadronisation, and DELPHES[32] (v3.4.2, GPL-3) using the default CMS detector card for the simulation of detector effects. For all samples, jets are reconstructed using the anti-$k_T$ jet algorithm[33] with a radius parameter of $R = 0.5$, a minimum transverse momentum of $p_T > 25$ GeV, and an absolute pseudo-rapidity of $|\eta| < 2.5$. To identify jets originating from $b$-quarks, a $b$-tagging algorithm with a $p_T$-dependent efficiency and mis-tagging rate is applied. Electrons and muons are selected with the same $p_T$ and $\eta$ requirements as for jets. No requirement is placed on the missing transverse momentum $E_T^{miss}$.

A large sample of simulated Standard Model (SM) $t\bar{t}$ production is generated with the top quark mass $m_t = 173$ GeV, and used for initial studies as well as the background model in the $Z'$ studies. It contains approximately 11M events after a basic event selection of exactly one electron or muon and at least four jets of which at least two are $b$-tagged. We further produce samples for the top mass analysis: ~0.2M events each at mass points of $m_t = 170, 171, 172, 173, 174, 175, 176$ GeV in order to build templates, as well as a training sample of ~12M total $t\bar{t}$ events produced in steps of 0.1 GeV to achieve an approximately flat $m_t$ distribution in the 166-176 GeV range. This sample is used for all $t\bar{t}$ reconstruction studies as well as the top mass analysis. A final sample with $m_t = 171.9$ GeV was produced to be used as pseudo-data for the top mass analysis. The value used was initially known by only one member of the team to avoid bias in the final mass extraction.

A sample of simulated SM $t\bar{t}H$ production, in which the Higgs boson decays to a pair of $b$-quarks, is generated to model the signal process for the $t\bar{t}H$ analysis. This sample has the same event selection as applied to the $t\bar{t}$ samples, with an additional requirement of at least six jets due to the additional presence of the Higgs boson. Training of SPA-NET is performed using 10M $t\bar{t}H$ events with at least two $b$-tagged jet, while the final measurement is performed using a distinct sample where 0.2M of 1.1M events satisfy the more stringent requirement of containing least four $b$-tagged jets. Training with the two-tag requirement achieved better overall performance than on the tighter four-tag selection, which follows the most recent ATLAS analyses in this channel[34]. The background in this analysis is dominated by $t\bar{t} + b\bar{b}$ production, which is modeled using a simulated sample in which the top and bottom pairs are explicitly included in the hard process generated by MADGRAPH_AMC@NLO; of the 1.3M events generated, 0.2M survive the event selection.

Finally, we produce Beyond the Standard Model (BSM) events containing a hypothetical $Z'$ boson that decays to a pair of top quarks, using the vPrimeNLO model[35] in MADGRAPH_AMC@NLO. One sample of 0.2M events is produced at each of $m_{Z'} = 500, 700, 900$ GeV to evaluate search sensitivity at a range of masses. A sample with an approximately flat $m_{Z'}$ distribution is generated for network training by generating events in 1 GeV steps between 400 and 1000 GeV. We match jets to the original decay products of the top quarks and Higgs bosons using an exclusive $\Delta R = (\sqrt{(\phi_j - \phi_d)^2 + (\eta_j - \eta_d)^2}) < 0.4$ requirement, such that only one decay product can be matched to each jet and vice versa, always taking the closest match. This method is adopted both in ATLAS and CMS analyses and allows a crisp definition of the correct assignments as well as categorization of events based upon which particles are reconstructable, as explained in the Particle Detector subsection.

We train all models on a single machine with a AMD EPYC 7502 CPU and 4 NVidia 3090 GPUs for training. Each model was trained for a period of 24 hours on this machine, as we have found that to be sufficient time for models to converge in training and validation loss. We use the same hyper-parameters derived in our previous work[16] as each event topology presented here may be interpreted as a variation of the same event topologies.

The data generated for this study is available in the our online repository (https://mlphysics.ics.uci.edu/data/2023_spanet/). The code used for training is available on github (https://github.com/Alexanders101/SPANet).

### Results and discussion
#### Reconstruction and regression performance
We present the reconstruction efficiency for SPA-NET in semi-leptonic $t\bar{t}$ and $t\bar{t}H(H \to b\bar{b})$ events, compared to the performance of the benchmark methods KLFitter and PDNN. Efficiencies are presented relative to all events in the generated sample, as well as relative to the subset of events in which all top quark (and Higgs boson in the case of $t\bar{t}H$) daughters are truth-matched to reconstructed jets, which we call Full Events. We also show efficiencies for each type of particle, with $t_H$ the hadronically decaying top quark, $t_L$ the leptonically decaying top quark, and $H$ the Higgs boson. We present the efficiencies in three bins of jet multiplicity as well as inclusively.

In Table 1, the efficiencies for accurate reconstruction of semi-leptonic $t\bar{t}$ events are shown. We find that SPA-NET outperforms both benchmark methods in all categories. The performance of KLFitter is substantially lower than the other two methods everywhere, reaching only 12% for full-event efficiency in full events with ≥6 jets. The PDNN performance is close to SPA-NET in low jet multiplicity events, but the gap grows as the number of jets in the event increases. This is expected due to the encoded symmetries in SPA-NET that allow it to more efficiently learn the high multiplicity, more complex events, as well as the additional permutations that must be considered by the PDNN. SPA-NET is further suited to higher-multiplicity events due to not suffering from the large run-time scaling of the permutation based approaches. Results for $t\bar{t}H(H \to b\bar{b})$ events, also presented in Table 1, show similar trends.

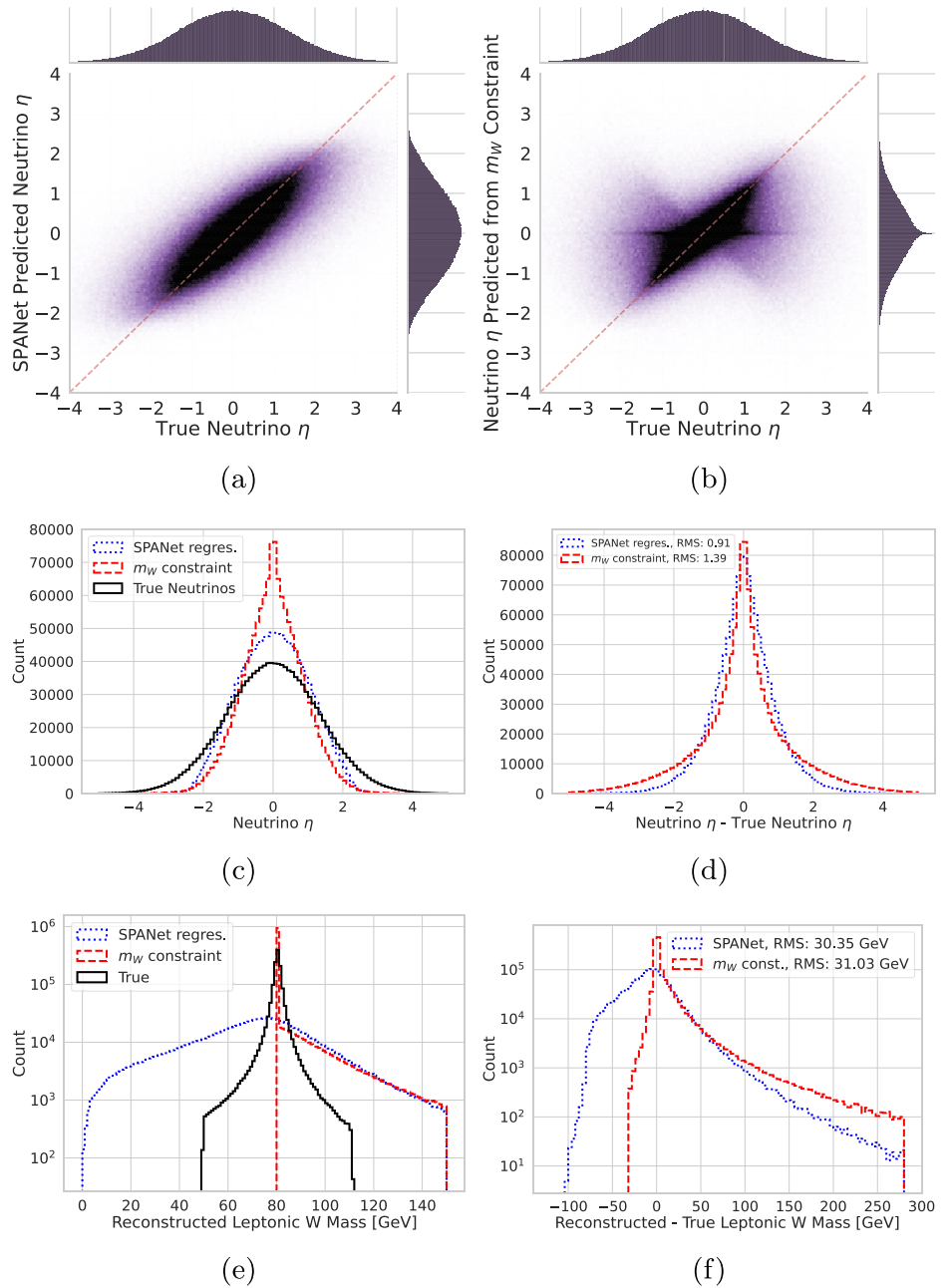

**Fig. 2 | Comparison of the regression of neutrino pseudo-rapidity ($\eta$) by SPA-NET with the benchmark $W$ boson mass constraint method. a, b** show the true value on the $x$-axis versus predicted values from the SPA-NET regression and $W$-mass constraint respectively on the $y$-axis, with the one-dimensional distributions shown outside the axes. **c** compares the neutrino $\eta$ from SPA-NET regression (blue dotted), $W$-mass constraint (red dashed), and the true distribution (black solid), with (**d**) showing the residuals between truth and SPA-NET regression (blue dotted) or $W$-mass constraint (red dashed). **e, f** show the same distributions, this time for the reconstructed leptonic $W$ boson mass.

**Regression performance.** In semi-leptonic $t\bar{t}$ decays, there is a missing degree of freedom due to the undetected neutrino. The transverse component and $\phi$ angle of the neutrino can be well-estimated from the missing transverse momentum in the event, but the longitudinal component (or equivalently, the neutrino $\eta$) cannot be similarly estimated at hadron colliders due to the unknown total initial momentum along the beam. A typical approach is to assume that the invariant mass of the combined lepton and neutrino four-vectors should be that of the $W$ boson, $m_W = 80.37$ GeV. This assumption leads to a quadratic formula that can lead to an ambiguity if the equation has either zero or two real solutions, and assumes on-shell $W$ bosons and perfect lepton and $E_T^{miss}$ reconstruction. When the equation has two real solutions, the one with the lower absolute value is adopted. If the solutions are complex, we take the real component.

SPA-NET has been extended to provide additional regression outputs, which can be used to directly estimate such missing components. In Fig. 2a, b, distributions of truth versus predicted neutrino $\eta$ show that the SPA-NET regression is more diagonal than the traditional $W$-mass-constraint method. Figure 2c, d shows the distributions and residuals of neutrino $\eta$, making it clear that SPA-NET regression has improved resolution of this quantity. However, Fig. 2e, f show that neither method is able to accurately reconstruct the $W$-mass distribution. This distribution is not regressed directly, but is calculated by combining the $E_T^{miss}$ and lepton information with the predicted value of $\eta$. The mass constraint method produces a large peak exactly at the $W$-mass as expected, with a large tail at high mass coming from events in which the quadratic solutions are complex. In contrast, the SPA-NET regression, which has no information on the expected value of the $W$-mass, has a similar shape above $m_W$, and a broad shoulder at lower values. It may thus be useful to refine the regression step to incorporate physics constraints, such as the $W$ boson mass, to help the network learn important, complex quantities such as this. Incorporating more advanced regression techniques, such as this or combining with alternative methods such as $\nu$-Flows[36,37], is left to future work.

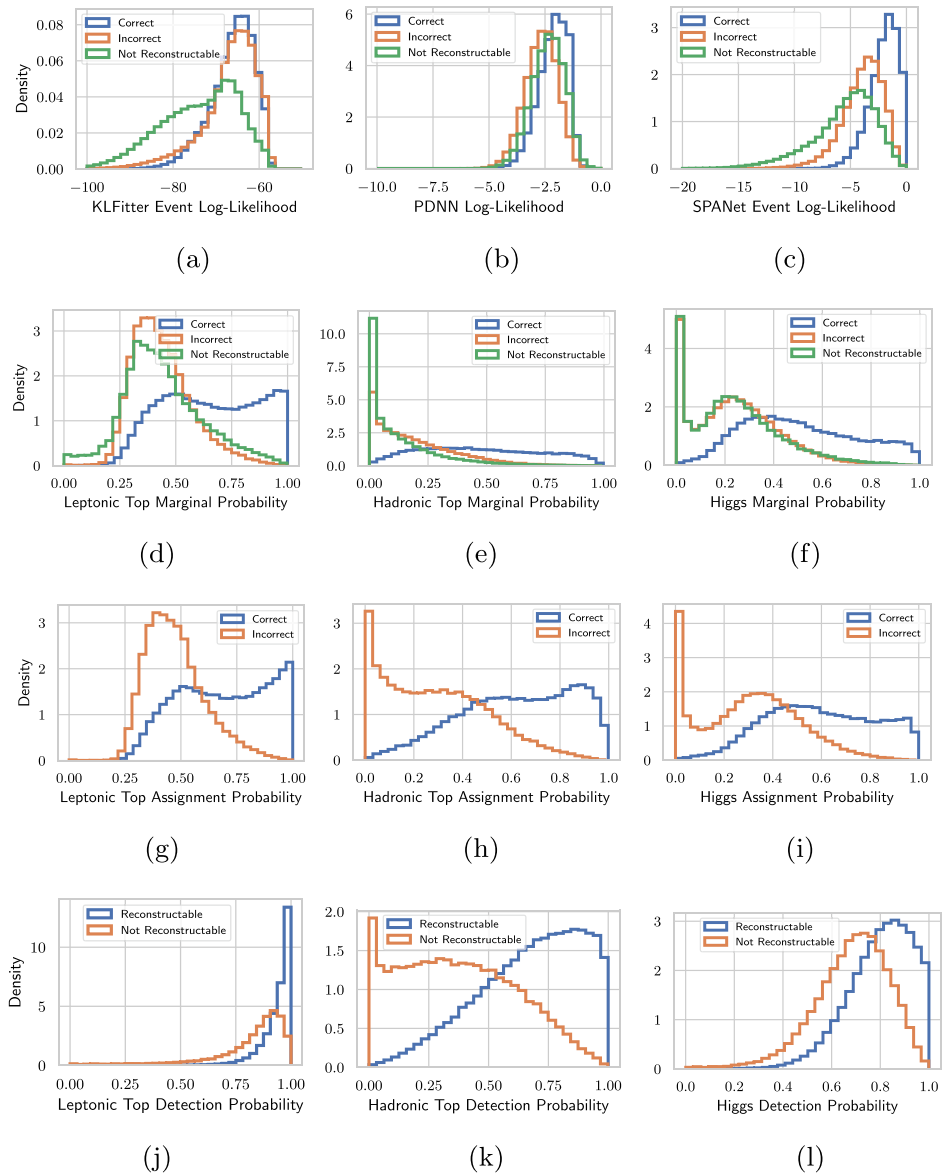

**Fig. 3 | Output distributions from SPA-NET and baseline methods.** The KLFitter likelihood is shown in (**a**), the Permutation Deep Neural Network (PDNN) log-likelihood in (**b**), and the SPA-NET event-level log-likelihood in (**c**), split by correctly reconstructed events (blue), incorrect events (orange), and non-reconstructable events (green). Further, the SPA-NET marginal probabilities for leptonic top, hadronic top, and Higgs are shown in (**d**–**f**, respectively, grouped in the same way. **g**–**i**) show the SPA-NET assignment probabilities, grouped by correct (blue) and incorrect (orange) events. Finally, the SPA-NET detection probabilities, split by reconstructable (blue) and non-reconstructable (orange), are shown in (**j**–**l**).

**Particle presence outputs.** The additional SPA-NET outputs, described in the Particle Detector subsection and shown in Fig. 3, can be very useful in analysis. The KLFitter, PDNN, and SPA-NET event-level likelihoods are shown in Fig. 3a–c. We note that the permutation methods only provide event-level scores for the entire assignment, and that the scores are highly overlapping with little separation between correctly and incorrectly reconstructed events. Figure 3d–f shows the SPA-NET per-particle marginal (pseudo-)probabilities, which are summed to calculate the event-level likelihood. The distributions of the assignment probability, separated by events, which SPA-NET has predicted correctly or incorrectly, are shown in Fig. 3g–i, and Fig. 3j–l shows the distribution of the detection probability split by whether the particle is reconstructable or not. All of the SPA-NET scores show clear separation between these categories, and this separation can be used in a variety of ways, such as to remove incomplete or incorrectly matched events via direct cuts, separate different types of events into different regions, or provide separation power as inputs to an additional multivariate analysis. The top quark mass and $Z'$ analyses both cut on these scores in order to remove incorrect/non-reconstructable events and improve signal-to-background ratio (S/B). In the $t\bar{t}H$ analysis, these are used as inputs to a Boosted Decision Tree (BDT) to classify signal and background, and are found to provide a large performance gain.

**Computational overhead.** Performance tests are performed on an AMD EPYC 7502 CPU with 128 threads and an NVidia RTX 3090 GPU. Including all pre-initialization steps, we evaluate the average run time for the three methods—KLFitter, PDNN, and SPA-NET—for both $t\bar{t}$ and $t\bar{t}H$ events. We find that KLFitter averages 24 (2) events per second on $t\bar{t}$ ($t\bar{t}H$). The PDNN averages 2626 (51) events per second when run on a CPU, and 3034 (101) events per second on a GPU, with the speed up from GPU hardware minimal due to the fact that permutation building dominates the computation time. In contrast, SPA-NET averages 705 (852) events per second on a CPU, and 4407 (3534) events per second on a GPU, showing reduced scaling with the more complex $t\bar{t}H$ events as expected. We therefore conclude that inference of SPA-NET should not be a bottleneck to analyses, as is often the case for methods like KLFitter. These numbers are summarized in table form in Supplementary Table 1.

**Ablation studies**
In this section, we present several studies designed to reveal what the networks have learned. We find that training is, in general, very robust, showing little dependence on details of inputs or hyperparameters. For example, training performance is unchanged within statistical uncertainties when representing particles using $\{M, p_T, \eta, \phi\}$ or $\{p_x, p_y, p_z, E\}$ 4-vector representations. Reconstruction performance varies by less than 1% if the training sample with a single top mass value is replaced by that with a flat mass spectrum.

In addition, we find that the performance of the network in testing depends on the kinematic range of the training samples in a sensible way. For example, the performance of the network on independent testing events varies with the top quark pair invariant mass, reflecting the mass distribution of the training sample. Figure 4 shows the testing performance versus top quark pair mass for networks trained on the full range of masses, or only events with invariant mass less than 600 GeV. The performance at higher mass is degraded when high-mass samples are not included in the training, as the nature of the task depends on the mass, which impacts the momentum and collimation of the decay products. Furthermore, the network performance is independent of the process (SM $t\bar{t}$ or BSM $Z'$) used to generate the training sample. The performance is reliable in the full range in which training data is present. It is noteworthy that the SM training still achieves similar performance up to ~1 TeV as the network trained on $Z'$ events, despite having fewer events at this value, indicating that the training distribution need not be completely flat so long as some examples are present in the full range.

To evaluate if the network is learning the natural symmetries of the data, we perform two further tests. The first is to investigate the azimuthal symmetry of the events, which we evaluate by applying the network to events that are randomly rotated in the $\phi$ plane and/or mirrored across the beam axis, which should have no impact on the nature of the reconstruction task. We find that in 41% of test events, the difference in the marginal probabilities is <1% and 84% of all events have a difference of less than 5%. This implies that the network approximately learns the inherent rotational and reflection symmetries of the task, without explicitly encoding this into the the network architecture. The full residual distributions are shown in Supplementary Fig. 1.

The impact of adding rotation invariance to the network has been evaluated by employing an explicitly invariant attention architecture which employs a matrix of relative Lorentz-covariant quantities between each pair of particles, similar to existing literature[18,38]. We focus specifically on the symmetry induced by rotations along the beam axis. We follow the covariant transformer architecture[18], and treat the $\phi$ and $\eta$ angles as covariant, and compute the difference between these angles for all pairs of jets in the event. The remaining features are treated as invariant and processed normally by the attention. Figure 5a shows that employing the invariant attention mechanism improves performance for small datasets, but does not lead to higher overall performance. This observation is consistent with the findings of existing literature[18,38]. The explicit invariance does bring visible improvement in training speed as seen in Fig. 5b. After fully training both networks on various training data sizes, we examine the training log and determine how many batches (gradient updates) were necessary before achieving maximal validation accuracy. We see that the invariant attention significantly reduces the number of updates needed to train the network. The trade-off of this regime is to make each network larger and more memory intensive, as the inputs must now be represented as pairwise matrices of features instead of simple vectors. Since the overall performance in the end is the same, and since we notice that a regular network already learns to approximate this invariance, we proceed using the traditional attention architecture, and this invariant network is not used for any further studies presented here.

**Search for $t\bar{t}H(H \to b\bar{b})$**
While the previous sections have detailed the per-event performance of SPA-NET, in the following sections we demonstrate its expected impact on flagship LHC physics measurements and searches.

The central challenge of measuring the cross-section for $t\bar{t}H$ production, in which the Higgs boson follows its dominant decay mode to a pair of $b$-quarks, is separating the $t\bar{t}H$ signal from the overwhelming $t\bar{t}+b\bar{b}$ background. Typically, machine learning algorithms such as deep neural networks or boosted decision trees are trained to distinguish signal and background using high-level event features[34,39]. Since the key kinematic difference between the signal and background is the presence of a Higgs boson, the performance of this separation is greatly dependent on the quality of the event reconstruction, where improvements by SPA-NET can make a significant impact on the final result.

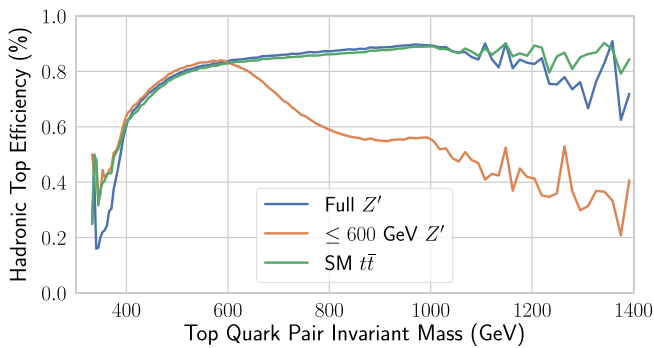

**Fig. 4 | Performance of the networks in testing data, as measured by hadronic top reconstruction efficiency, as a function of the top quark pair invariant mass.** Shown is the performance for three networks with distinct training samples: $Z' \to t\bar{t}$ events with the full range of invariant masses (blue), $Z' \to t\bar{t}$ events with masses <600 GeV (orange), and SM $t\bar{t}$ with the full range of invariant masses (green).

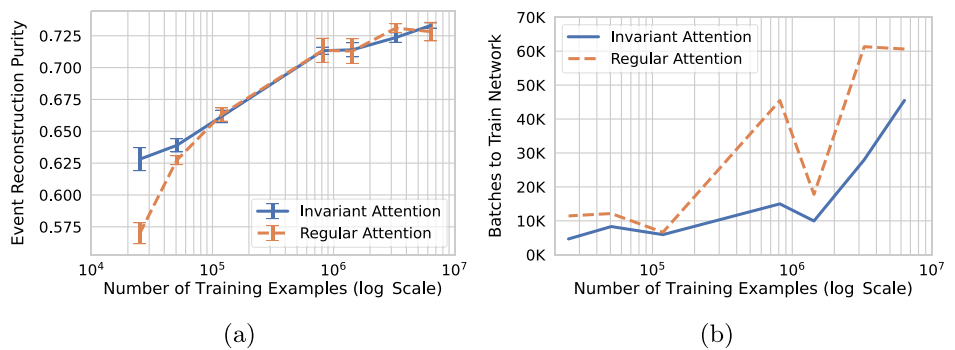

**Fig. 5 | A comparison between the regular transformer and the explicitly invariant transformer as a function of training dataset size.** Shown are (**a**) reconstruction purity and (**b**) training speed, with the regular transformer shown in dashed orange and the explicitly invariant transformer[18] in solid blue. The uncertainty bars in (**a**) show the variation in reconstruction purity across 16 separate trainings at each dataset size.

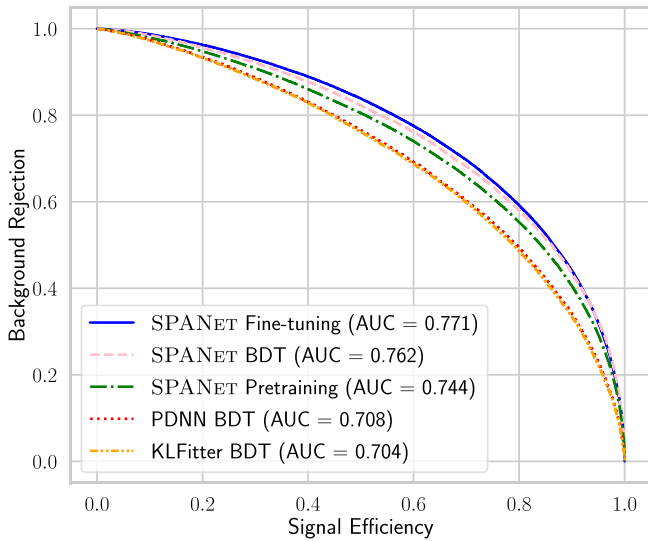

Fig. 6 | Receiver operating curve for networks trained to distinguish $t\bar{t}H$ from the major background $t\bar{t} + b\bar{b}$. Shown is signal efficiency versus background rejection for several SPA-NET based set ups—SPA-NET fine-tuning (solid blue), SPA-NET +Boosted Decision Tree (BDT) (dash-dot pink), and SPA-NET pretraining (dash-dot green)—as well as BDTs based on outputs of traditional reconstruction techniques Permutation Deep Neural Network (PDNN) (dotted red) and KLFitter (dot-dash yellow).

**Reconstruction and background rejection.** Event reconstruction is performed with SPA-NET, KLFitter, and a PDNN. The reconstruction efficiency for each of these methods is shown in Table 1, where it is already clear that SPA-NET outperforms both of the baseline methods.

The reconstructed quantities and likelihood or network scores are then used to train a classifier to distinguish between signal and background. The full input list is shown in Supplementary Table 2, with most variable definitions taken from the latest ATLAS result[34]. A BDT is trained for each reconstruction algorithm with the same input definitions and hyperparameters using the XGBoost package[40]. Tests using a BDT trained on lower-level information, i.e., the four-vectors of the predicted lepton and jet assignments, found significantly weaker performance than these high-level BDTs. We also compare the performance of the BDTs to two different SPA-NET outputs that are trained to separate signal and background. The first, which we call SPA-NET Pretraining, is an additional output head of the primary SPA-NET network, which has the objective of separating signal and background events. The second, which we call SPA-NET Fine-tuning, uses the same embeddings and central transformer as the former method, but the signal versus background classification head is trained in a separate second step after the initial training is complete. In this way, the network is able to first learn the optimal embedding of signal events, and utilize this embedding as the inputs to a dedicated signal vs background network. We have implemented in the SPA-NET package an option to output directly the embeddings from the network such that they can be used in this or other ways by the end user.

The receiver operating curve for the various classification networks is shown in Fig. 6. The best separation performance comes from the fine-tuned SPA-NET model, as expected. The BDT with kinematic variables reconstructed with the SPA-NET jet-parton assignment (SPA-NET+BDT setup) is next, followed by the purely pre-trained model. All of these substantially outperform both the KLFitter+BDT and PDNN+BDT baselines.

**Impact on sensitivity.** To estimate the impact of significantly improved signal-background separation from SPA-NET reconstruction, we perform an Asimov fit to the network output distributions with the pyhf package[41,42]. The signal is normalized to the SM cross-section of 0.507 pb[43] and corrected for the branching fraction and selection

**Table 2 | Expected large hadron collider (LHC) Run 2 (Run 3) sensitivity to $t\bar{t}H$ as measured in a parameterized detector model described in the text**

|  | Signal significance | Upper cross-section limit [pb] | Upper signal strength limit |
|---|---|---|---|
| KLFitter BDT | 2.4σ (4.1σ) | 0.426 (0.248) | 0.840 (0.489) |
| PDNN BDT | 2.4σ (4.1σ) | 0.421 (0.246) | 0.831 (0.486) |
| SPA-NET BDT | 3.0σ (5.2σ) | 0.340 (0.196) | 0.671 (0.387) |
| SPA-NET pre-training | 2.7σ (4.8σ) | 0.371 (0.214) | 0.732 (0.423) |
| SPA-NET fine-tuning | 3.1σ (5.7σ) | 0.332 (0.179) | 0.655 (0.353) |

Shown is the expected statistical significance of the measurement as well as expected upper limits on cross-section and signal strength using the output of classification networks trained on the products of various reconstruction algorithms. The signal strength is defined as the ratio of the Standard Model prediction to the measured cross-section.

efficiency of our sample. The dominant $t\bar{t} + b\bar{b}$ background is normalized similarly, using the cross-section calculated by MAD-GRAPH_AMC@NLO of 0.666 pb. We further multiply the background cross-section by a factor of 1.5, in line with measurements from ATLAS[34] and CMS[39] that found this background to be larger than the SM prediction, rounded up to account also for the LO→NLO cross-section enhancement. We neglect the sub-leading backgrounds. The distributions are binned according to the AutoBin feature[44] preferred by ATLAS in order to ensure no bias is introduced between the different methods due to the choice of binning. Results normalized to 140 fb$^{-1}$, the luminosity of Run 2 of the LHC, using 5 bins and assuming an overall systematic uncertainty of 10% are presented in Table 2. The numbers in the parentheses in Table 2 are results of an LHC Run 3 analysis normalized to 300 fb$^{-1}$ of data using 8 bins with an overall systematic uncertainty assumption of 7%. Although the Run 3 center-of-mass energy of the LHC is $\sqrt{s} = 13.6$ TeV, all results presented assume $\sqrt{s} = 13$ TeV for simplicity.

In both scenarios, the sensitivity tracks the signal-background separation performance shown in Fig. 6, with SPA-NET fine-tuning achieving the greatest statistical power. Neither of the benchmark methods is able to reach the 3σ statistical significance threshold in the Run 2 analysis, while both SPA-NET+BDT and fine-tuning reach this mark. Similarly, these methods both reach the crucial 5σ threshold normally associated with discovery, with the benchmark methods at only roughly 4σ.

SPA-NET thus provides a significant expected improvement over the benchmark methods. While the full LHC analysis will require a more complete treatment, including significant systematic uncertainties due to the choice of event generators, previous studies have demonstrated minimal dependence to such systematic uncertainties[16].

**Top mass measurement**
The top quark mass $m_t$ is a fundamental parameter of the Standard Model that can only be determined via experimental measurement. These measurements are critical inputs to global electroweak fits[45], and $m_t$ even has implications for the stability of the Higgs vacuum potential, which has cosmological consequences[46,47]. Precision measurements of the top quark mass are thus one of the most important pieces of the experimental program of the LHC, with the most recent results reaching sub-GeV precision[48–50]. We demonstrate in this section the improvement enabled by the use of SPA-NET in a template-based top mass extraction.

We perform a two-dimensional fit to the invariant mass distributions of the hadronic top quark and $W$ boson as reconstructed by each method, using the basic preselection described in the Datasets and Training sub-section. We further truncate the mass distributions to $120 \leq m_t \leq 230$ GeV and $40 \leq m_W \leq 120$ GeV. The fraction of events with correct or incorrect predictions for the top quark jets has a strong impact on the resolution with which the mass can be extracted. Better reconstruction should thus improve the overall sensitivity to the top quark mass.

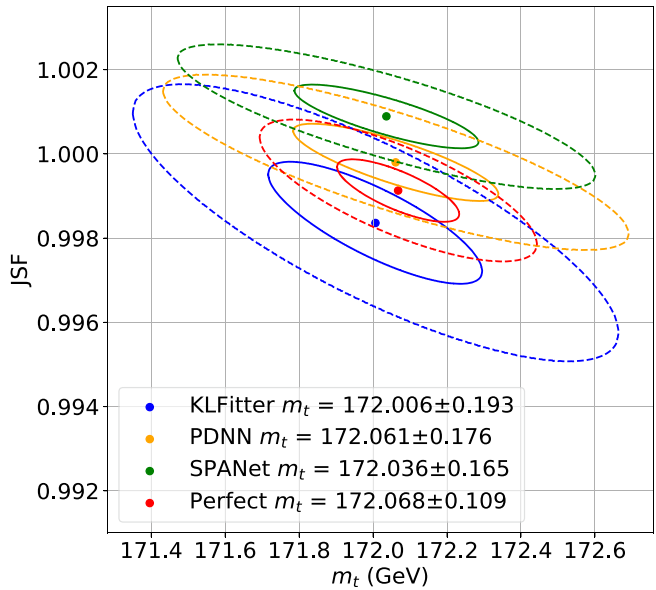

**Fig. 7 | Expected best-fit top quark mass ($m_t$) and jet scale factor (JSF) from a template-based Asimov fit.** Shown are results for the KLFitter (blue), permutation deep neural network (PDNN) (yellow), SPA-NET (green), and an idealized perfect reconstruction (red). Also shown are $1\sigma$ (solid) and $3\sigma$ (dashed) uncertainty ellipses.

**Table 3 | Expected global significance for a $Z'$ signal with an integrated luminosity of 140 (300) fb$^{-1}$, for several choices of $Z'$ mass and reconstruction algorithms**

|  | KLFitter | PDNN | SPA-NET | SPA-NET w/$\eta^\nu$ |
|---|---|---|---|---|
| $m_{Z'} = 500$ GeV | 1.2σ (2.5σ) | 1.8σ (3.5σ) | 2.8σ (5.5σ) | 2.7σ (5.4σ) |
| $m_{Z'} = 700$ GeV | 1.6σ (3.3σ) | 2.5σ (4.9σ) | 3.1σ (6.1σ) | 2.9σ (5.7σ) |
| $m_{Z'} = 900$ GeV | 1.9σ (3.9σ) | 2.8σ (5.5σ) | 4.3σ (8.5σ) | 4.1σ (8.2σ) |

Incorporation of the W-mass information in the 2D fit allows for a simultaneous constraint on the jet energy scale uncertainty, often a leading contribution to the total uncertainty, by also fitting a global jet scale factor (JSF) to be applied to the $p_T$ of each jet. Further, events that do not contain a fully reconstructable top quark are removed by cutting on the various scores from each method. KLFitter events are required to have a log-likelihood score >−70, PDNN events must have a network score of >0.12, and SPA-NET events must have a marginal probability of >0.23, optimized in each case to minimize the uncertainty on the extracted top mass. We additionally compare each method to an idealized perfect reconstruction method, in which all unmatched events are removed, and the truth-matched reconstruction is used for all events. The perfect-matched method provides an indication of the hypothetical limit of improvement achievable through better event reconstruction. In all cases, we neglect background from other processes, since these backgrounds tend to be on the order of a few percent[25], and would be further suppressed by the network score cuts.

The top quark mass and JSF are extracted using a template fit from Monte Carlo samples which have top quark masses in 1 GeV intervals between 170 and 176 GeV. Templates are constructed for varying mass and JSF hypotheses for both the top and W boson mass distributions. These templates are built separately for each of the correct, incorrect, and unmatched event categories as the sum of a Gaussian and a Landau distribution, with five free parameters: the mean $\mu$ and the width $\sigma$ of each, as well as the relative fraction $f$. We found an approximately linear relation between the template parameters as a function of the top quark mass and JSF, allowing for linear interpolation between the mass points. Finally, we validate the mass extracted by a template fit in hypothetical similar experiments and find a small bias, for which we derive a correction.

The impact of various reconstruction techniques can be best measured by the resulting uncertainty on the top quark mass and JSF. Figure 7 shows the expected uncertainty ellipses for a dataset with luminosity of 140 fb$^{-1}$ and assuming a JSF variation of ±4%. The final uncertainty on the top mass is 0.193 GeV for KLFitter, 0.176 GeV for PDNN, and 0.165 GeV for SPA-NET. This indicates a 15% improvement in top quark mass uncertainty when using SPA-NET compared to the benchmark methods. The idealized reconstruction technique achieves an uncertainty of 0.109 GeV, demonstrating how much room for improvement remains. The dominant contribution to the gap between the perfect and SPA-NET reconstruction comes from the perfect removal of all unmatched events.

### Search for $Z' \to t\bar{t}$

Many BSM theories hypothesize additional heavy particles which may decay to $t\bar{t}$ pairs, such as heavy Higgs bosons or new gauge bosons ($Z'$). We investigate a generic search for such a $Z'$ particle, for which accurate reconstruction of the $t\bar{t}$ mass peak over the SM background plays a crucial role. We compare the performance of the benchmark reconstruction methods to that of various SPA-NET configurations by assessing the ability to discover a $Z'$ signal.

An important aspect is the selection of training data, due to the unknown mass of the $Z'$, which strongly affects the kinematics of the $t\bar{t}$ system. To avoid introducing bias into the network, the training sample is devised to be approximately flat in $m_{t\bar{t}}$. The network training was otherwise identical to that described for the SM $t\bar{t}$ network, and performance on SM $t\bar{t}$ events was approximately the same in the mass range covered by both samples.

The basic $t\bar{t}$ selection described in the Dataset and Training subsection is applied, and all events are reconstructed as described earlier in order to calculate the $t\bar{t}$ invariant mass, $m_{t\bar{t}}$. The mass resolution of a hypothetical resonance can often be improved by removing poorly- or partially-reconstructed events. In the context of the algorithms under comparison, this corresponds to a requirement on the KLFitter likelihood or network output scores. The threshold is chosen to optimize the analysis with each algorithm, leading to a significant reduction of the SM $t\bar{t}$ background when using the PDNN and SPA-NET. For SPA-NET we require a marginal probability of >0.078, and for PDNN we require a score of >0.43. For KLFitter, no cut is applied, as no improvement was found. More details on these cuts and the effect on the background distributions are shown in Supplementary Figs. 2 and 3 in Supplementary Note 1.

**Impact on sensitivity.** We use the pyhf[41,42] package to extract the $Z'$ signal and assess statistical sensitivity.

The expected results for a Run 2 analysis, normalized to 140 fb$^{-1}$ with 20 GeV bins and a systematic uncertainty of 10%, are shown in Table 3. The discovery significance is improved by SPA-NET compared to the benchmark methods for all masses considered. For example, for a $Z'$ of mass 700 GeV the limit improves from 1.6σ using KLFitter to 3.1σ using SPA-NET.

The expected sensitivity in a Run 3 dataset with the integrated luminosity of 300 fb$^{-1}$ is computed with an optimistic systematic uncertainty of 5% as also shown in Table 3. For all the three benchmark signals, discovery significance exceeds 5σ using SPA-NET, while for the baseline methods only the high mass point for the PDNN reaches this threshold. At a $Z'$ mass of 500 GeV, KLFitter does not reach the 3σ evidence threshold, while SPA-NET is able to make a discovery. It is noteworthy that the neutrino regression does not lead to an improvement on the final sensitivity, despite showing improved resolution compared to the baseline mass constraint method. This is due to the effect on the background shape, which similarly improves in this case.

Improved reconstruction with SPA-NET can therefore greatly boost particle discovery potential. This finding should extend to other hypothetical resonances such as heavy Higgs bosons, $W'$ bosons, or SUSY particles as well as non-$t\bar{t}$ final states such as di-Higgs, di-boson, $tb$ or any other in which reconstruction is crucial and challenging.

## Conclusions

This paper describes significant extensions and improvements to SPA-NET, a complete package for event reconstruction and classification for high-energy physics experiments. We have demonstrated the application of our method to three flagship LHC physics measurements or searches, covering the full breadth of the LHC program; a precision measurement of a crucial SM parameter, a search for a rare SM process, and a search for a hypothetical new particle. In each case, the use of SPA-NET provides large improvements over benchmark methods. We have further presented studies exploring what the networks learn, demonstrating the ability to learn the inherent symmetries of the data and strong robustness to training conditions. SPA-NET is the most efficient, high-performing method for multi-object event reconstruction to date and holds great promise for helping unlock the power of the LHC dataset.

## Data availability

Our data is available in an online repository.

## Code availability

Our code is available on github (https://github.com/Alexanders101/SPANet).

## Acknowledgements
We would like to thank Ta-Wei Ho for assistance in generating some of the samples used in this paper. D.W. and M.F. are supported by DOE grant DE-SC0009920. The work of A.S. and P.B. in part supported by ARO grant 76649-CS to P.B. H.O. and Y.L. are supported by NSFC under contract no. 12075060, and SCH is supported by NSF under Grant no. 2110963.

## Author contributions
Michael Fenton: conception, direction, supervision of all students, manuscript preparation. Alexander Shmakov: development, implementation, and training of SPA-NET and PDNN, manuscript preparation. Hideki Okawa: MC production, $Z'$ analysis lead, manuscript preparation, supervision of Y. Li. Yuji Li: $t\bar{t}H$ analysis lead Ko-Yang Hsiao: top mass analysis lead Shih-Chieh Hsu: supervision of K-Y Hsiao Daniel Whiteson: manuscript preparation, supervision of A. Shmakov. Pierre Baldi: manuscript editing, machine learning developments, supervision of A. Shmakov.

## Competing interests
The authors declare no competing interests.

## Additional information
**Supplementary information** The online version contains supplementary material available at https://doi.org/10.1038/s42005-024-01627-4.

**Correspondence** and requests for materials should be addressed to Michael James Fenton or Alexander Shmakov.

**Peer review information** *Communications Physics* thanks Daniel Murnane and the other, anonymous, reviewer(s) for their contribution to the peer review of this work. A peer review file is available

**Reprints and permissions information** is available at http://www.nature.com/reprints

**Publisher's note** Springer Nature remains neutral with regard to jurisdictional claims in published maps and institutional affiliations.

# Supplementary Table 1: Average Inference Time

|            | $t\bar{t}$              | $t\bar{t}H$             |
|------------|-------------------------|-------------------------|
| KLFitter   | 24 events per second    | 2 events per second     |
| PDNN CPU   | 2626 events per second  | 51 events per second    |
| PDNN GPU   | 3034 events per second  | 101 events per second   |
| SPANet CPU | 705 events per second   | 852 events per second   |
| SPANet GPU | 4407 events per second  | 3534 events per second  |

**Supplementary Table 1**: Average run time for jet-parton assignment inference from various algorithms on $t\bar{t}$ and $t\bar{t}H$ events.

# Supplementary Figure 1: Azimuthal Symmetry

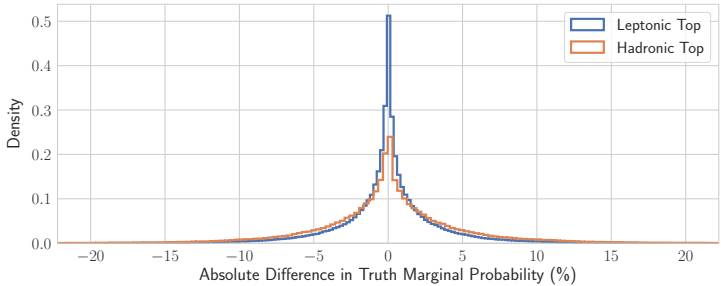

**Supplementary Figure 1**: Difference in marginal probability between a nominal event and a $\phi$-rotated and/or mirrored version of the same event, for the leptonic (blue) and hadronic (orange) top quarks.

# Supplementary Table 2: BDT inputs for $t\bar{t}H$

| Variable | Definition |
|---|---|
| General kinematic variables | |
| $\Delta R_{bb}^{\text{avg}}$ | Average $\Delta R$ for all $b$-jet pairs |
| $\Delta R_{bb}^{\max\ p_\text{T}}$ | $\Delta R$ between the two $b$-jets with the largest vector sum $p_\text{T}$ |
| $\Delta \eta_{jj}^{\max}$ | Maximum $\Delta\eta$ between any jet pairs |
| $H_\text{T}^{\text{had}}$ | Scalar sum of jet $p_\text{T}$ |
| $m_{bb}^{\min\ \Delta R}$ | Mass of two $b$-jets with the smallest $\Delta R$ |
| $m_{jj}^{\min\ \Delta R}$ | Mass of any jet pair with the smallest $\Delta R$ |
| $N_{bb}^{\text{Higgs 30}}$ | Number of $b$-jet pairs with invariant mass within 30 GeV of the Higgs mass |
| $\Delta R_{l,bb}^{\min}$ | Smallest $\Delta R$ between the lepton and the combination of the two b-jets |
| Variables with Higgs boson and top quark reconstruction | |
| $m_{bb}^{\text{Higgs}}$ | Mass of the Higgs boson candidate |
| $m_{H,b_{\text{lep top}}}$ | Mass of the Higgs boson candidate and $b$-jet from leptonic top quark candidate |
| $\Delta R_{bb}^{\text{Higgs}}$ | $\Delta R$ between $b$-jets from the Higgs boson candidate |
| $\Delta R_{H,t\bar{t}}$ | $\Delta R$ between Higgs boson candidate and $t\bar{t}$ candidate system |
| $\Delta R_{H,\text{leptop}}$ | $\Delta R$ between Higgs boson candidate and leptonic top quark candidate |
| $\Delta R_{H,b_{\text{hadtop}}}$ | $\Delta R$ between Higgs boson candidate and $b$-jet from hadronic top candidate decay |
| Scores from jet-parton assignment | |
| LHD | Log-likelihood discriminant from the KLFitter |
| $A_{\text{PDNN}}$ | Assignment score from Permutation DNN |
| $A_{\text{higgs}}$ | Assignment probability of Higgs boson target from SPA-NET |
| $D_{\text{higgs}}$ | Detection probability of Higgs boson target from SPA-NET |
| $M_{\text{higgs}}$ | Marginal probability of Higgs boson target from SPA-NET |
| $E_{\text{higgs}}$ | Assignment entropy of Higgs boson target from SPA-NET |
| $A_{\text{leptop}}$ | Assignment probability of leptonic top quark candidate from SPA-NET |
| $D_{\text{leptop}}$ | Detection probability of leptonic top quark candidate from SPA-NET |
| $M_{\text{leptop}}$ | Marginal probability of leptonic top quark candidate from SPA-NET |
| $E_{\text{leptop}}$ | Assignment entropy of leptonic top quark candidate from SPA-NET |
| $A_{\text{hadtop}}$ | Assignment probability of hadronic top quark candidate from SPA-NET |
| $D_{\text{hadtop}}$ | Detection probability of hadronic top quark candidate from SPA-NET |
| $M_{\text{hadtop}}$ | Marginal probability of hadronic top quark candidate from SPA-NET |
| $E_{\text{hadtop}}$ | Assignment entropy of hadronic top quark candidate from SPA-NET |

**Supplementary Table 2**: Input variables to the classification Boosted Decision Tree (BDT) in our $t\bar{t}H$ analysis. Only the score(s) from one jet-assignment algorithm under consideration, i.e. KLFitter, Permutation Deep Neural Network (PDNN), or SPA-NET, is used in the BDT and limit setting.

# Supplementary Note 1: Quality Cuts for $Z'$

In typical $Z'$ searches [1, 2], it is usual to apply a quality cut on the reconstruction algorithm, such as by eliminating events below a given threshold in the KLFitter likelihood or network output scores. By removing a large number of partial events or those with incorrect assignments, the mass resolution can be further improved. We tune this cut for each method to find optimal S/B for each mass point and chose the average of the optimal thresholds across the three considered mass points to be applied in our analysis. Supplementary Figure 2 shows the receiver operating curve for this scan, with the optimal point for each mass denoted by a filled red star (green circle) for SPA-NET (PDNN) and the selected cut value denoted by an unfilled marker. It is found that the KLFitter likelihood actually has no impact for $Z'$ of mass 500 GeV and actually removes more signal than the background for the 700 and 900 GeV benchmarks. Thus, we do not apply a cut on the KLFitter likelihood in our analysis. These cuts lead to a significant reduction of the SM $t\bar{t}$ background when using the PDNN and SPA-NET.

Another important consideration in $Z'$ analyses is the sculpting of the background distributions induced by the reconstruction method. To assess this, Supplementary Figure 3 shows the $t\bar{t}$ invariant mass distributions for each method compared to the truth distribution. The numbers in the legend are the Earth Movers Distance (EMD) metric, a measure of the difference between the truth and reconstructed distributions. Figure 3 (left) shows the distributions before the quality cuts, and demonstrate that each method has a degree of sculpting towards lower masses, with KLFitter sculpting the least and PDNN the most. Supplementary Figure 3 (right) shows the same distributions for each method after the quality cuts, with this time a different truth distribution per method as each contains a different set of events. The sculpting is still visible when using the PDNN, with the EMD slightly increased, suggesting slightly increased sculpting. In comparison, the SPA-NET, the invariant mass distributions both with and without the neutrino $\eta$ regression is almost compatible with the truth distribution across the full mass range. The $\eta$ regression leads to a slightly reduced EMD indicating reduced sculpting.

# Supplementary References

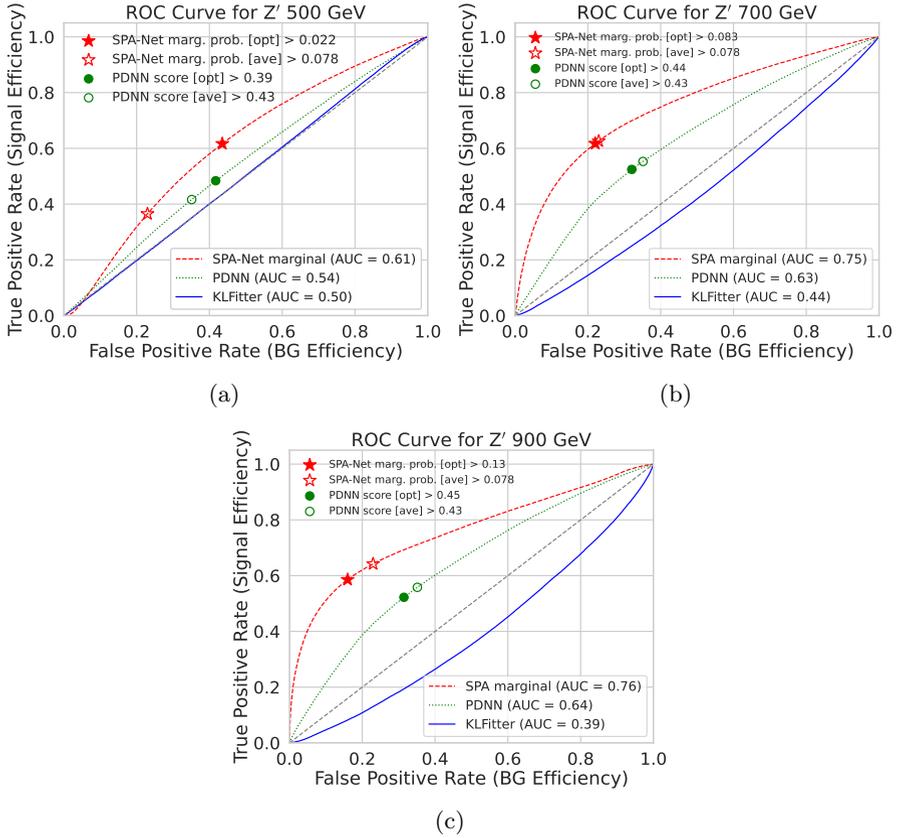

**Supplementary Figure 2**: Receiver operating characteristic curves presented for the KLFitter likelihood (solid blue line), Permutation Deep Neural Network (PDNN) score (green dotted line / green circles) and SPA-NET marginal probability (dashed red line / red stars) for (a) 500 GeV, (b) 700 GeV and (c) 900 GeV signal benchmarks. The optimal threshold for each signal (solid markers) and mean of the three optimal thresholds (open markers) are shown.

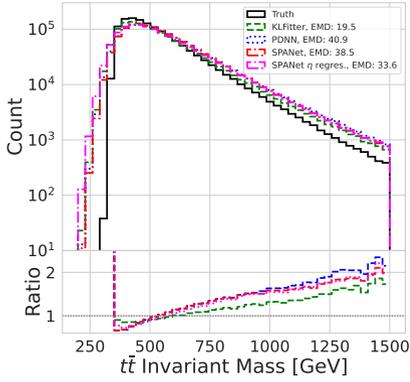 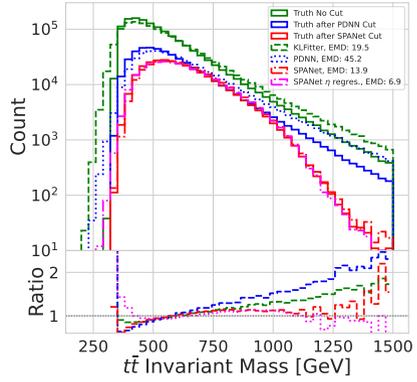

(a)           (b)

**Supplementary Figure 3**: Invariant mass of the $t\bar{t}$ system for various jet-parton assignment algorithms, (a) without applying any score cut, and (b) after applying optimized score cuts. In panel (a), truth is shown as a black solid line, KLFitter with a green dashed line, Permutation Deep Neural Network (PDNN) with a blue dotted line, and SPA-NET without (with) $\nu^\eta$ regression with a red (pink) dot-dashed line. In panel (b), the truth distribution after score cuts is shown as solid lines in the corresponding color of each method as defined in panel (a).